\documentclass[12pt,preprint]{aastex}
\usepackage{epsfig}
\usepackage{graphicx}

\def\ltsima{$\; \buildrel < \over \sim \;$}
\def\lsim{\lower.5ex\hbox{\ltsima}}
\def\gtsima{$\; \buildrel > \over \sim \;$}
\def\gsim{\lower.5ex\hbox{\gtsima}}

\shorttitle{The HST Supernova Survey}
\shortauthors{Dahlen et al.}

\begin{document}


\title{The Extended HST Supernova Survey: The Rate of SNe Ia at $z>1.4$~Remains Low}


\author{Tomas Dahlen\altaffilmark{1}, 
Louis-Gregory Strolger\altaffilmark{1,2},
and Adam G. Riess\altaffilmark{1,3}
}
\email{dahlen@stsci.edu}

\altaffiltext{1}{Space Telescope Science Institute, 3700 San Martin Drive, Baltimore, MD 21218}
\altaffiltext{2}{Department of Physics and Astronomy, Western Kentucky University, Bowling Green, KY 42101}
\altaffiltext{3}{Department of Physics and Astronomy, Johns Hopkins University, 3400 North Charles Street, Baltimore, MD 21218}



\begin{abstract}
We use the $HST$~ACS imaging of the two GOODS fields during Cycles 11, 12, and 13 to derive the Type Ia supernova rate in four redshift intervals in the range $0.2<z<1.8$. Compared to our previous results based on Cycle 11 observations only, we have increased the coverage of the search by a factor 2.7, {\it making the total area searched equivalent to one square degree}. The sample now consists of 56 Type Ia supernovae. The rates we derive are consistent with our results based on the Cycle 11 observations. In particular, the few detected supernovae at $z>1.4$~supports our previous result that there is a drop in the Type Ia supernova rate at high redshift, suggesting a long time delay between the formation of the progenitor star and the explosion of the supernova. If described by a simple one-parameter model, we find a characteristic delay time of 2--3 Gyr. However, a number of recent results based on e.g., low redshift supernova samples and supernova host galaxy properties suggest that the supernova delay time distribution is bimodal. In this model, a major fraction of the Type Ia supernova rate is 'prompt' and follows the star formation rate, while a smaller fraction of the rate has a long delay time, making this channel proportional to mass. While our results are fully consistent with the bimodal model at low redshifts, the low rate we find at $z>1.4$~appears to contradict these results. Models that corrects for star formation hidden by dust may explain at least part of the differences. Here we discuss this possibility together with other ways to reconcile data and models.
\end{abstract}


\keywords{
supernovae: general -- surveys
}


\section{Introduction}
The use of Type Ia supernovae (hereafter SNe) as standardizable candles has led to surprising and intriguing results when it comes to the expansion rate of the universe, however, it is still a most unsatisfactory situation that the specifics of the progenitor scenarios that lead to the SN explosion are mostly unknown. Two general scenarios of how a white dwarf (WD) accretes matter to reach the Chandrasekhar mass and thereafter explode are often discussed. In the single degenerate (SD) scenario, the WD accretes mass from a companion sub-giant or giant star, while the double degenerate (DD) scenario involves the merger of two WDs. The different progenitor scenarios are expected to involve different delay time distributions, i.e., the time between the formation of the progenitor star and SN explosion (e.g., Yungelson \& Livio 2000; Greggio 2005).

The rate of Type Ia SNe should mimic the cosmic star formation rate (SFR), albeit shifted to lower redshift by the delay time distribution set by the progenitor scenario. In Dahlen et al. (2004) and Strolger et al. (2004), we used a sample of 23 Type Ia SNe detected with $HST$~ACS in the GOODS fields to derive the SN rate (SNR) to $z\sim$1.6. Using the observed rate and a simple one-component model characterized by a single delay-time, $\tau$, we found with 95\%~confidence $\tau > 2$~Gyr. The main observational feature driving $\tau$~to a high value, was the deficit of detected SNe at $z > 1.4$.

While this was the first attempt to derive the SNR at $z>1$~to set constraints on the delay time, different approaches to address the question have also been taken. The single delay-time model offers a simple parametrization of the relation between SFR and SNR (Madau et al. 1998; Dahlen \& Fransson 1999), however, it is plausible that there are different channels, characterized by different delay-times, contributing to the Type Ia SNR, making the single-model an oversimplified description of the data. Mannucci et al. (2005) calculated the Type Ia SNR as a function of stellar mass in galaxies, and found that overall rate could be described as the sum of a 'young' or 'prompt' component proportional to the ongoing SFR and an 'old' component proportional to the stellar mass.  Similar results are found by Sullivan et al. (2006), who use results from the Supernova Legacy Survey (SNLS) to derive the Type Ia SN rate as a function of host galaxy SFR and mass. Their results also support a scenario where at least one component of the SNR is closely related to the instant SFR. Additional observations that support that a two-component model is plausible includes, e.g., Branch \& van den Bergh (1993), who found differences in the spectra of Type Ia SNe found in early-type and late-type galaxies, respectively. Furthermore, a number of results suggest differences in peak magnitude distributions between SNe occurring in different environments, in particular, under-luminous Type Ia SNe are mostly found in early type galaxies, while the brightest Type Ia SNe are preferentially found in late-type galaxies (Filippenko 1989; Della Valle \& Panagia 1992; Hamuy et al. 1995; 2000; Reindl et al. 2005). If SNe with different characteristics (e.g., peak magnitudes) occur in different environments ('old' vs. 'young' galaxies), then it is conceivable that different explosion mechanisms are involved, e.g., one channel that produces the faint SNe in 'old' galaxies, and another that produces the bright SNe in 'young' galaxies. Furthermore, Scannapieco \&~Bildsten (2005) show that a two-component model with a large contribution of prompt SNe can help explain the Fe content of the inter-cluster medium in clusters of galaxies, as well as the [O/Fe] abundance ratios in the Galaxy.

While the SNR we derived in Dahlen et al. (2004) is consistent with other measurements and models with a large 'prompt' component at $z<1$, there is an apparent discrepancy at $z \gsim 1.4$~between the long delay-time ($\tau=2-4$~Gyr) derived from the one-component fit to the GOODS data (Strolger et al. 2004) and models supporting a large 'prompt' component, e.g., Scannapieco \&~Bildsten (2005; hereafter SB05) and Neill et al. (2006; hereafter N06). The model presented by Mannucci et al. (2007; hereafter M07), which includes a larger fraction of the 'old' component and a different treatment of dust extinction, appears to fit data better.

In this paper, we make use of data from the continued monitoring of the GOODS fields within the PANS program to increase the temporal coverage of the survey by a factor $\sim 2.7$, allowing us to better constrain the high redshift SNR. In particular, we investigate the proposed decline in SNR at $z>1.4$.

We organize this paper as follows: In \S 2 we describe our data and in \S 3 we present our results. A discussion follows in \S 4, while we in \S 5 focus on the bimodality of Type Is SNe. In \S 6 we briefly discuss results at lower redshifts. Finally, conclusions and a summary are given in \S 7.

\section{Data}
In this investigation, we use the original GOODS data, consisting of one reference image and four search images in each of GOODS-North and GOODS-South obtained during $HST$~Cycle 11 (Riess et al. 2004), together with the PANS continued monitoring of during $HST$~Cycles 12-13. This added nine search images in GOODS-North and five search images in GOODS-S (Riess et al. 2007). In total, we have therefore searched 22 fields, each consisting of 15 pointings in the F850LP filters using $HST$~ACS. With each pointing being $\sim$11 sq arcmin, {\it we have in total searched an area of $\sim$one square degree with an average time separation of 45 days.} In compliment to the F850LP search filter, observations with the F775W filter (and F606W filter in Cycle 11) added contemporaneous color information at all epochs. We note here that $HST$~has a number of unique features making it more reliable than recent ground-based surveys for detecting and confirming high redshift ($z>1$) SNe. This includes high signal-to-noise detections made possible through low background, in particular in the reddest filters, and a high spatial resolution which is important for distinguishing SNe from variable nuclei in their host galaxies. Furthermore, during this campaign we conducted a rolling search where the same two fields were observed repeatedly every 40-50 days, allowing multiple, homogeneous measurements of the SN lightcurves without interruption (due to weather) as well as the construction of an extremely deep reference image for the latter epochs. Finally, $HST$ with the g800l grism on ACS WFC has the capability to achieve high signal-to-noise spectra to $\sim$1 micron to identify the redshift and type of SN Ia up to $z \sim 1.4$. This ability provided reliable identifications for the majority of our SNe Ia at the highest-redshift bin of our survey (Riess et al. 2007), a feature unmatched in other surveys at this redshift range. Details on the observations and reduction techniques used are given in Giavalisco et al. (2004) and Strolger et. al (2004).

\section{Results}
A total of 56 Type Ia SNe were discovered during the full program, of which 23 belong to the original GOODS set, while 33 were discovered in the follow up program. The SNe are classified as ``gold'' (33 SNe), ``silver'' (9 SNe), or ``bronze'' (14 SNe) depending on the confidence of the classification. A gold classification requires either a high S/N spectrum of the SN (available for 30 of 33), or a spectroscopic redshift of the host galaxy together with three additional requirements: 1) the host is of elliptical/early type, 2) the peak rest-frame $U-V$~color is consistent with a SN Ia, and 3) the light-curve shape is consistent with a SN Ia (3 of 33 are based on these criteria). If two of the three criteria above are met, then the SN is considered a silver Type Ia. Finally, if only the light-curve shape is consistent with a Type Ia, then a bronze confidence level is assigned. While the bronze classification implies that the SNe are most probably Type Ia, we can not rule out that a few may be interlopers, i.e., misidentified core collapse SNe. This classification is further discussed in Strolger et al. (2004). Spectroscopic redshift of either the SN or the host galaxy is available for 48 of the SNe. For the remaining eight, we rely on photometric redshifts. At high redshift ($z>1$), our sample consists of 23 Type Ia. Of these, we have spectroscopic redshift of the SN or the host galaxy for 21 objects. Divided by classification, the high redshift sample consists of 15 gold, 5 silver, and 3 bronze Type Ia SNe. The SNe used in this investigation are described in detail in Riess et al. (2004; 2007) and Strolger et al. (2004, 2008 in prep.). For a full catalog of GOODS/PANS SNe with type classification and positions, visit the GOODS/PANS catalog web page \footnote{http://zwicky.wku.edu/}. Additional information on this sample, including redshifts, will be published in Strolger et al. (2008, in prep.).

We derive the SNR as described in detail in Dahlen et al. (2004). In short, we use Monte Carlo simulations to estimate the predicted number of detected SN for a search with the same observational setup as GOODS/PANS, e.g., field-of-view, detection limits and temporal spacing between observations. Input to the simulations is a given 'intrinsic' SN rate in units number of SNe per year per Mpc$^3$. Taking into account the SN peak magnitude luminosity function, K-corrections, light-curve shape, light-curve stretch and extinction of the SN light due to dust in the host galaxy (see \S 4.2 for details), we calculate the expected number of detected SNe in each redshift interval under consideration. We thereafter adjust the input rate so that the expected number of detected SNe matches the actual number of SNe found in the search. To account for the relative large redshift errors for the SNe with only photometric redshifts available, we use the redshift probability distribution derived from the photometric redshift fitting when distributing these SNe in bins. Therefore, a single SNe may contribute with fractional parts in more than one bin.

\begin{figure}
\epsscale{0.7}
\plotone{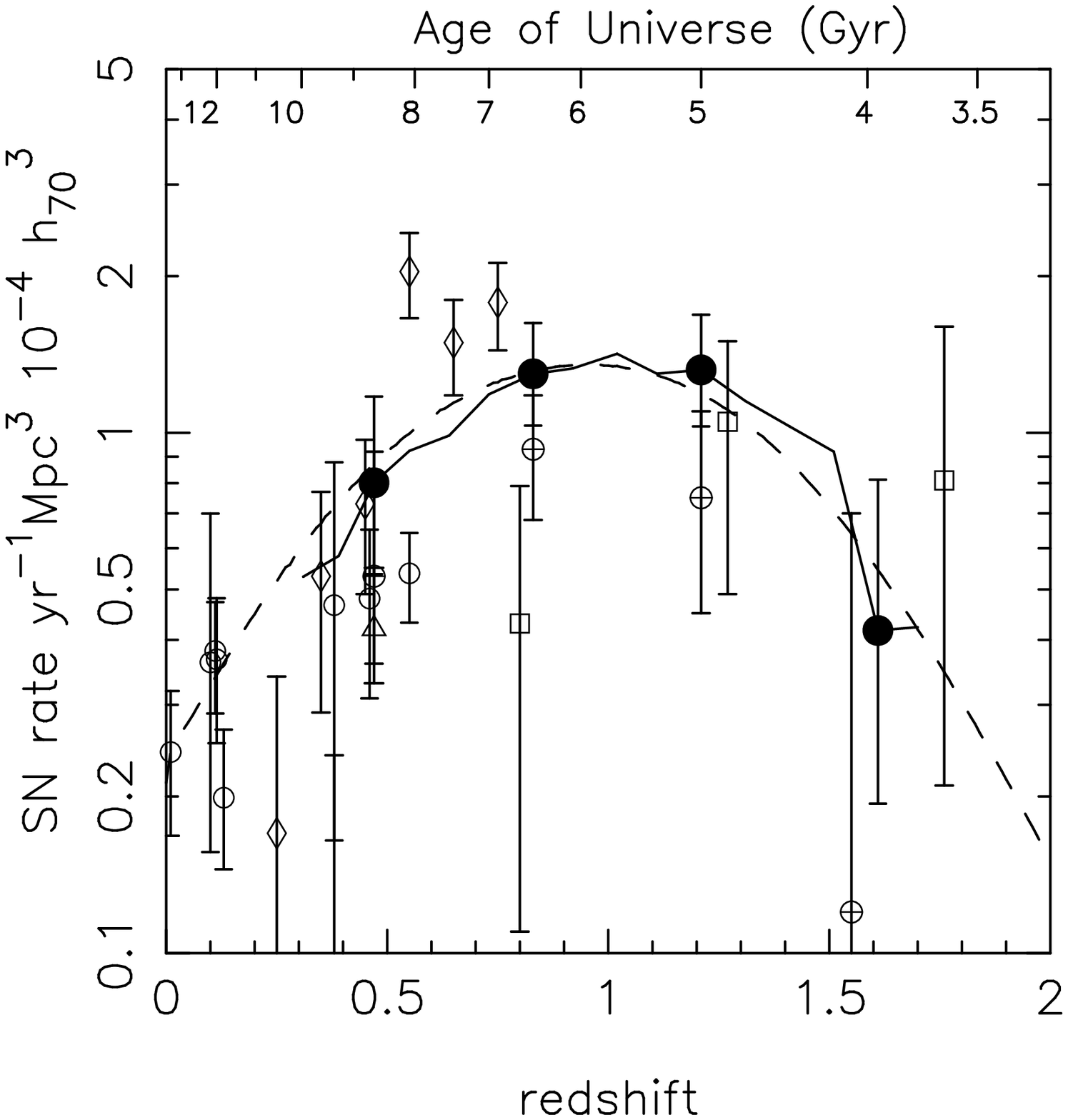}
\figcaption[f1.eps]{Type Ia supernova rates. Black dots show the rates derived in this investigation. The solid line shows the SNR between $z=0.3$~and $z=1.7$~calculated using a window function with width $\Delta z=0.4$. Dashed line shows the best-fitting one component delay-time model with $\tau=3.4$~Gyr. Open circles show rates from the literature and are taken from Cappellaro et al. (1999), Hardin et al. (2000), Strolger (2003), Reiss et al. (2000), Blanc (2004), Pain et al. (1996), Tonry et al. (2003), and Pain et al. (2002). Rates from Barris \& Tonry (2006) are shown with diamonds, rates from Neill et al. (2006) with triangle, rates from Poznanski et al. (2007) are shown with squares and rates from Kuznetsova et al. (2007) are shown as crossed circles. Error-bars represent 1$\sigma$ statistical errors, except for the measurements of Poznanski et al. (2007), which also include uncertainty in type determination.
\label{fig1}}
\end{figure}

In Table \ref{tab1}~and Figure \ref{fig1}, we show our resulting rates in four redshift bins $0.2<z<0.6$, $0.6<z<1.0$, $1.0<z<1.4$, and $1.4<z<1.8$. In the Figure, we show our results as black dots. In Table \ref{tab1}, we also give the 'raw' number of Type Ia SNe in each bin together with the number of SNe after taking into account the redshift distribution of the SNe with photometric redshift only. Error-bars for our points in Figure \ref{fig1} only include statistical errors, however, in Table \ref{tab1}, we also give systematic errors. These include uncertainties in SN classifications. To estimate the maximum contamination from core collapse SNe, we assume that all bronze Type Ia SNe have been misclassified. To estimate the maximum number of SNe that we may have excluded due to misclassification, we assume that all bronze CC SNe detected in the survey are instead Type Ia SNe. The systematic effects also include uncertainties in the Type Ia SN luminosity function, K-corrections and extinction corrections. Details on how the systematic errors are derived are given in Dahlen et al. (2004).

To investigate how the adopted binning affects results, we also calculate rates using a window function of width $\Delta z=0.4$, which we shift from redshift bin $0<z<0.4$~to $1.5<z<1.9$. Results are shown with the solid line in Figure \ref{fig1}.

It is clear from Figure ~\ref{fig1} that the steep decrease in the SNR at $z>1.4$~found in Dahlen et al. (2004) is still present in our extended data set. We also note that the suggested peak in the Ia rate at $z\sim$0.8 from the old data is now not as pronounced, rather the rate seems fairly flat over the range $0.8<z<1.2$. 
\section{Discussion}
In Figure \ref{fig1}, we also include a compilation of rates taken from the literature; Cappellaro et al. (1999), Hardin et al. (2000), Strolger (2003), Reiss et al. (2000), Blanc et al. (2004), Pain et al. (1996; 2002), Tonry et al. (2003), Neill et al. (2006), Barris \& Tonry (2006), Kuznetsova et al. (2007) and Poznanski et al. (2007). Considering the relatively large errors on these measurements, there is a fairly good agreement between rates determined at similar redshifts. To make a quantitative measurement of how well the data from different surveys agree, we fit to a third order polynomial using all data points in Figure \ref{fig1}. This results in a reduced chi-square $\chi^2/\nu$=2.6, reflecting some of the existing discrepancies found in the rates derived by different authors. In particular, the rates from Barris \& Tonry at $z\sim 0.6$~are a factor $\sim$2 higher than other measurements at comparable redshifts. Also, our rate at $z\sim 0.47$~is almost twice the rate measured recently by Neill et al. (2006) using the SNLS data, and three times the rates of Poznanski et al. (2007) at $z\sim 0.8$. Furthermore, the rate at $z>1.4$~presented in Poznanski et al. is about twice the rate found here. We discuss these apparent discrepancies in \S 5 and \S 6.

\begin{figure}
\epsscale{0.7}
\plotone{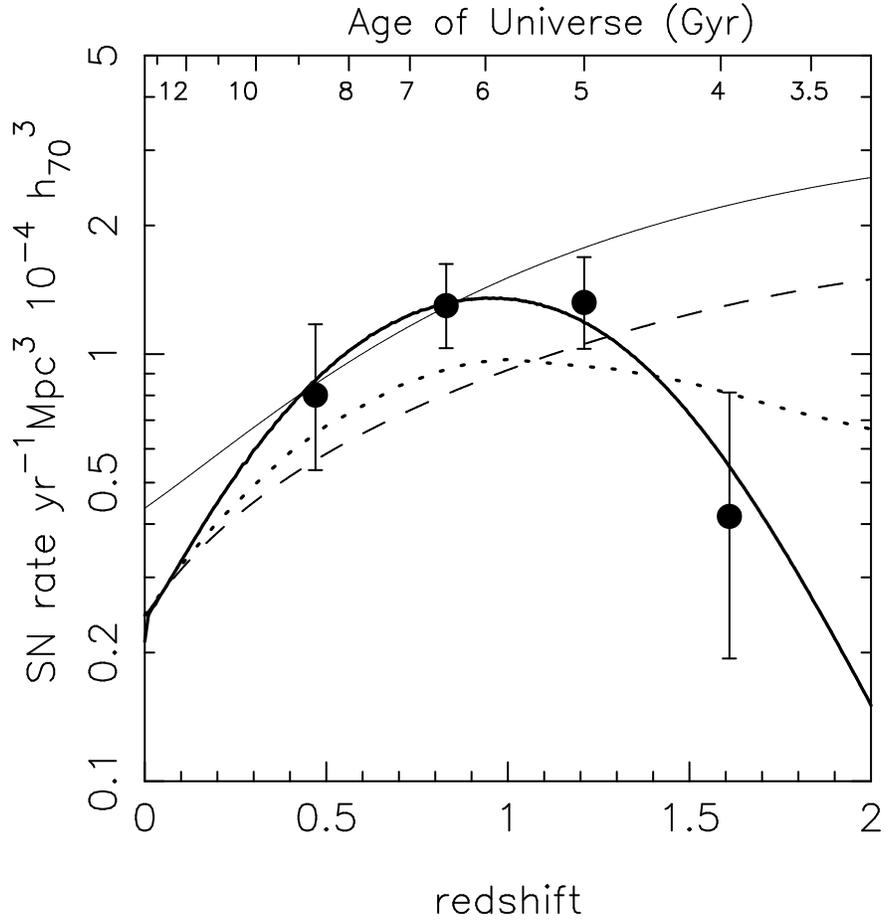}
\figcaption[f2.eps]{Type Ia supernova rates derived in this investigation are shown as filled circles. The thick solid line shows the best fitting one-component model assuming a delay-time $\tau=$ 3.4 Gyr. Also plotted are the two-component models from Scannapieco \& Bildsten (2005; thin solid line), Neill et al. (2006; dashed line) and Mannucci et al. (2007; dotted line). 
\label{fig2}}
\end{figure}

The best-fitting one-component model to the GOODS data has a Gaussian delay
time distribution with parameters $\tau=3.4$~Gyr and $\sigma=0.2\tau$~(for a definition of the delay-time distributions, see Strolger et al. 2004). This fit is shown in Figure \ref{fig1} as the dotted line and in Figure \ref{fig2} as the thick solid line. The fit is largely consistent with the one-sigma errors of a number of rate measurements over the wide redshift range. However, as previously indicated, there are several constraints that disfavor exceptionally long delay-times. Most prominently, the increased SNR in low-redshift star forming galaxies suggests that at least some SN Ia progenitors prefer channels which are more closely connected to ongoing star formation. Although the Strolger et al. (2004) one-component delay-time model cannot be ruled out (with high confidence) by the rate measurements alone, the low-redshift observations seemingly contradicts this simple one-parameter model which would decouple the SNR from the ongoing SFR. 

Also plotted in Figure \ref{fig2} are the two-component models from SB05 (thin solid line) and N06 (dashed line). These have the major contribution from a component that is directly proportional to the ongoing SFR, and a smaller component proportional to the old stellar component (i.e., the amount of accumulated mass). While these models are fairly consistent with measurements at $z\lsim~1.2$, they differ significantly at $z\gsim~1.4$~($>3\sigma$). The low measured rate at $z\sim~1.6$ is therefore difficult to reconcile with the two-component models of SB05 and N06. M07 calculate the expected SNR from a two-component model using a different approach compared to SB05 and N06. M07 include in the total SFR also the SFR that is hidden to optical searches due to dust extinction, i.e., contributions from faint IR galaxies (FIRGs), luminous IR galaxies (LIRGs) and ultra-luminous IR galaxies (ULIRGs). Taking into account that also a large fraction of the Type Ia SNe should be hidden in these galaxies, they find an observable Type Ia SNR that turns over at $z\sim1$. We show this model as the dotted line in Figure \ref{fig2}. This model shows a better agreement at the highest redshift point, while slightly under-predicting the rate at mid redshifts. Below, we further discuss these different models together with other effects that may affect the comparisons between observations and models, including statistical fluctuations, clustering variance, detection efficiency, dust extinction, and evolving WD explosion efficiency.

\subsection{Statistical fluctuations and clustering variance}
Adopting the two-component model rates from SB05 and N06, we find that 16 and 10 SNe should be detected in our highest redshift bin using the detection efficiency adopted here. This is significantly more than the three SNe detected. Using statistical error, we find that the probability is $\sim 10^{-4}$~of finding the predicted number from SB05, and $\sim 2*10^{-2}$~of finding the number predicted by N06. Therefore, it seems unlikely that the difference is due to statistical fluctuations alone, especially compared to the rate in SB05. Compared to N06, the difference is not ruled out with high confidence, but the difference is high enough that alternative explanations have to be investigated. Both SB05 and N06 models predict that $>$90\% of the SNR comes from the 'prompt' component at $z\sim$1.4. The M07 model on the other hand, only assumes $\sim$50\%, which together with the different assumptions about SFR and dust extinction leads to a lower number of expected high redshift SNe. Thus, in the highest redshift bin, the M07 model predicts 6 SNe which is consistent with our measurement within one sigma. Fitting the models to all data points using statistical errors results in reduced chi-square $\chi^2/\nu$=7.6, 4.0, and 2.1 for the SB05, N06, and M07 models, respectively. Again, this shows that deviations are fairly large compared to expected statistical variations, with exception for the M07 model, which is marginally consistent with data.

Clustering variance could affect the derived high redshift rates if there is a significant under-density in the observed fields. Using two widely separates field, GOODS-S and GOODS-N, should help to decrease these effects. Also, the star formation rate density measured in GOODS-S (Dahlen et al. 2007) to $z\sim~2$ is consistent with measurements in the same redshift range obtained for other fields. This implies that we should not be missing part of the prompt component of the SNR due to an under-density of star forming galaxies. Since the number counts of galaxies in GOODS-N and GOODS-S are comparable (GOODS-N slightly higher), we also don't expect the star formation rate in GOODS-N to be under-dense and thus affecting the number of detected SNe.

\subsection{Dust extinction}
\begin{figure}
\epsscale{0.7}
\plotone{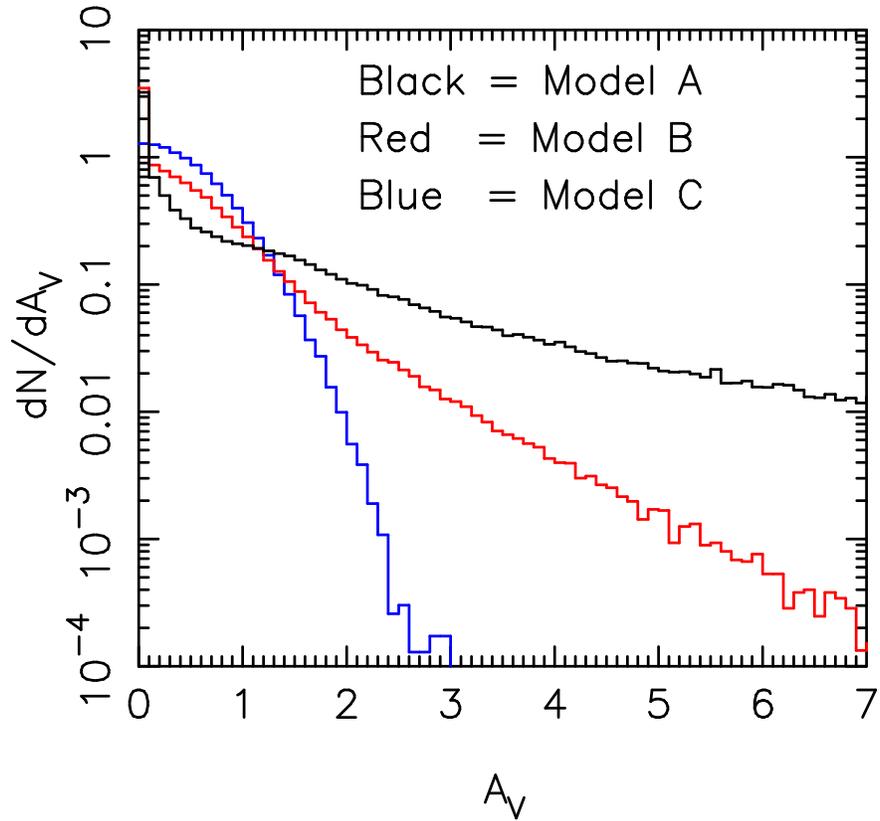}
\figcaption[f3.eps]{Distribution of V-band extinction, $A_V$, for the three different models tested here. Model A (black line) is the one we use in this paper to correct rates for extinction. This distribution is derived using results in Hatano et al. (1994). Model B (red line) is constructed to reproduce the distribution in Riello \& Patat (2005), while Model C (blue line) shows the distribution in Neill et al. (2006). 
\label{fig3}}
\end{figure}

Assumptions on the amount of dust extinction in the SN host galaxies also affects the derived SN rates. For example, a high extinction prior in the models results in a high derived SN rate, and vice verse. In this investigation, we have followed the prescription in Hatano et al. (1998) to calculate the extinction distribution. We hereafter refer to this as Model A. We note that the Hatano et al. model was derived for estimating extinction in local hosts and may therefore not be representative at high redshifts. To investigate the effect of assumed extinction distribution models, we have also considered the models in Riello \& Patat (2005) and N06. The models in Riello \& Patat (2005) use the same MC technique as Hatano et al. (1998), but include a more detailed approach aimed at being applicable also to non-local hosts. In particular, they include a tail of high extinctions for highly inclined host galaxies. Here we follow the approach in N06, who show that at low inclinations ($i<60^{\circ}$), the Riello \& Patat models can be approximated by a positive valued Gaussian (described below). For high inclinations, we assign a random extinction value according to E(B--V)=-ln$\Re/\lambda_{E(B-V)}$, where $\Re$~is a random number between 0 and 1 and $\lambda_{E(B-V)}$~describes how fast the high extinction tail falls of. Here we use $\lambda_{E(B-V)}$=5 at inclinations $60^{\circ}<i<75^{\circ}$~and $\lambda_{E(B-V)}$= 3 at $75^{\circ}<i<90^{\circ}$. Finally, to make the distribution of $A_V$~consistent with Figure 2 in Riello \& Patat (2005), we adjust the model so that $\sim$35\%~of the distribution falls within $0.0<A_V<0.1$. We call the model based on the Riello \& Patat distributions Model B. In N06, the distribution of E(B--V) is given by a positive valued Gaussian with $\sigma_{E(B-V)}$=0.2, $R_V$=3.1 and an extinction law following Cardelli et al. (1989). We include this parameterization as Model C. The three different distributions are shown in Figure \ref{fig3}. The main differences between the distributions are that Model C lacks the high extinction tail (at $A_V\gsim 2$) and has significantly less power at the low extinction end, i.e, this model has $\sim$13\% with $0.0<A_V<0.1$, compared to $\sim$35\% for both Model A and Model B. As a consequence, Model C has higher power at intermediate absorptions ($0.1<A_V<1.2$). Model A and Model B are more similar, both with a significant peak at the lowest absorptions, but with Model A having higher power at the high extinction tail ($A_V>1.2$), whereas Model B is dominating at intermediate absorptions. 

\begin{figure}
\epsscale{0.9}
\plotone{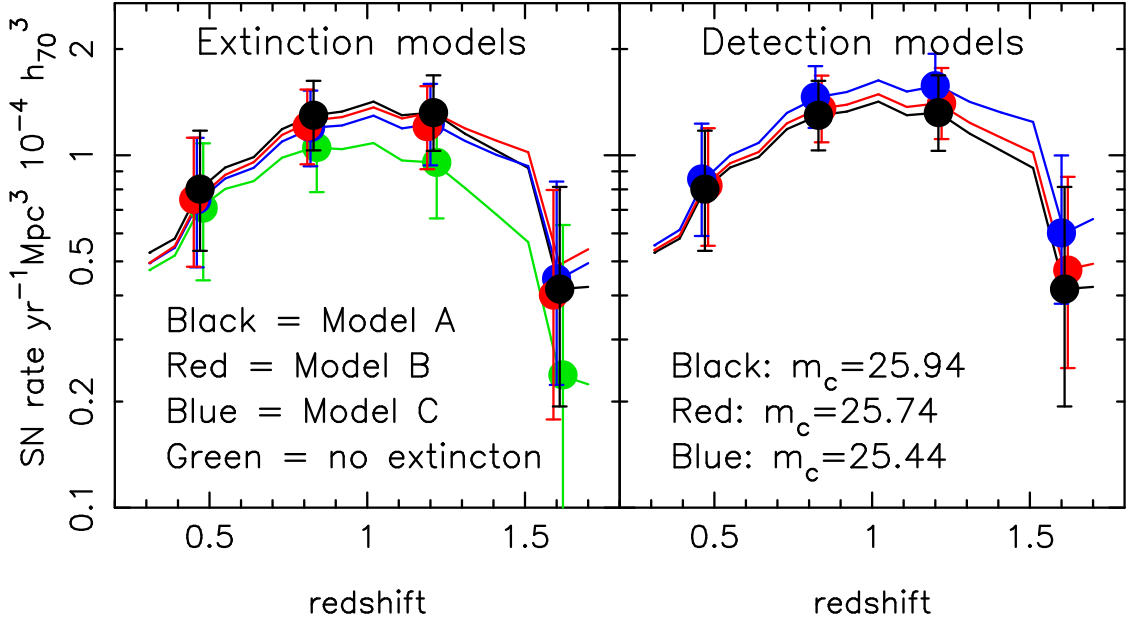}
\figcaption[f4.eps]{Left panel: Type Ia supernova rates derived from the GOODS/PANS program. Black dots shows rates derived using Model A, our nominal host galaxy dust extinction model. Red dots show Model B which is based on results from Riello \& Patat (2005). Model C is shown with blue dots and assumes a positive valued Gaussian distribution in E(B-V) with $\sigma_{E(B-V)}=0.2$~taken from Neill et al. (2006). For comparison, we also shows rates assuming no host galaxy extinction as green dots. Right panel: Black dots shows rates derived with nominal detection efficiency, i.e., the 50\% detection efficiency is reached at $m_c=25.94$. Red dots show rates assuming a magnitude cutoff 0.2 mag brighter than the nominal value, i.e, $m_c=25.74$, while blue dots show the case of a 0.5 mag brighter cutoff ($m_c=25.44$). 
\label{fig4}}
\end{figure}

In the left panel of Figure \ref{fig4}, we show our rates using the nominal assumption about host galaxy extinction distribution, i.e., Model A, together with the alternative distributions derived from Model B and Model C. We also show results in the case where no dust extinction is included. Assuming Model B, we expect a lower average extinction (see Figure \ref{fig3}) and therefore lower derived rates. We find that the Model B produces rates that are $\lsim$10\% lower compared to Model A. In particular, the rates in the highest redshift bin decreases by $\sim$4\% assuming Model B. Model C affects the derived rates in a different way so that the rates decrease at low redshift and increase at high redshift. For the highest bin, we find that the derived rate increases by $\sim$7\%~assuming Model C. It is surprising to note that the resultant SNR shows this little variation with assumed extinction model. If fact, the only notable exception is for no assumed host galaxy extinction, most notably in the highest redshift bin at $z=1.6$.  The low rate of SNe Ia in the high redshift bin is therefore a robust conclusion for the viable regime of assumed host galaxy dust models.

Besides the distribution of extinction within the host galaxies studied here, a change in the mean extinction in host galaxies with redshift may also affect results. We discuss this more in \S 5.
\subsection{Detection efficiency}
The detection efficiency of a survey is a crucial ingredient when deriving SN rates using both MC simulations and the control time method. Assuming that a survey reaches deeper than is the actual case, will underestimate derived rates. In the GOODS survey we characterized the detection efficiency with a function
\begin{equation}
\epsilon(\Delta m)=\frac{1}{1+e^{(\Delta m -m_c)/S}}
\end{equation}
where $\Delta m$~is the detection magnitude (difference between detection and reference epoch) and $m_c$~gives the magnitude where the detection efficiency drops to 50\%. The parameter $S$~gives the slope of the roll off in the detection efficiency. For the GOODS part of the survey we used two methods to determine the detection efficiency (Strolger et al. 2004). First, fake SN were added to the actual images that were being searched and the detection efficiency was derived from how many of the input SNe were recovered as a function of magnitude. In the second method, we used MC simulations and automated detection algorithms to derive the efficiency. Combining results we derived $m_c$=25.94 and $S$=0.38 (see Fig. 9 in Strolger et al. 2004). In the PANS part of the survey, the exposure time of the monitoring images was reduced by 20\% (from 2000s to 1600s). However, in the GOODS part, we used single epoch reference frames in the subtraction, while we in the latter PANS part of the survey added multiple templates to the reference images. This increased their depth, adding less noise signal when subtracting the reference image from the detection image. To quantify this, we use the ACS Exposure Time Calculator and calculate the S/N for a faint source ($m_z=25.9$) making two different assumptions. First, we calculate the signal for the GOODS exposure time (2000s) and the associated noise for a single epoch. To get the total noise in the subtracted image, we add back-ground, dark, and readout noise from detection image and reference image in quadrature to the source noise from the detection image. This gives us the estimated S/N in the subtracted image. Second, for the PANS search, we make the same calculation but with the signal given by a 1600s exposure and where we decrease the noise in the reference image by a factor sqrt(5), since at least five images were added when creating the reference image. We thereafter calculate the corresponding S/N in the subtracted image. We find that the PANS part of the survey results in similar (slightly higher) S/N compared to the GOODS part. This suggests that we reach at least as deep in the PANS part of the survey and should not be missing SNe despite the somewhat shorter exposure time. We can therefore use the same detection efficiency for the whole program.

\begin{figure}
\epsscale{0.9}
\plotone{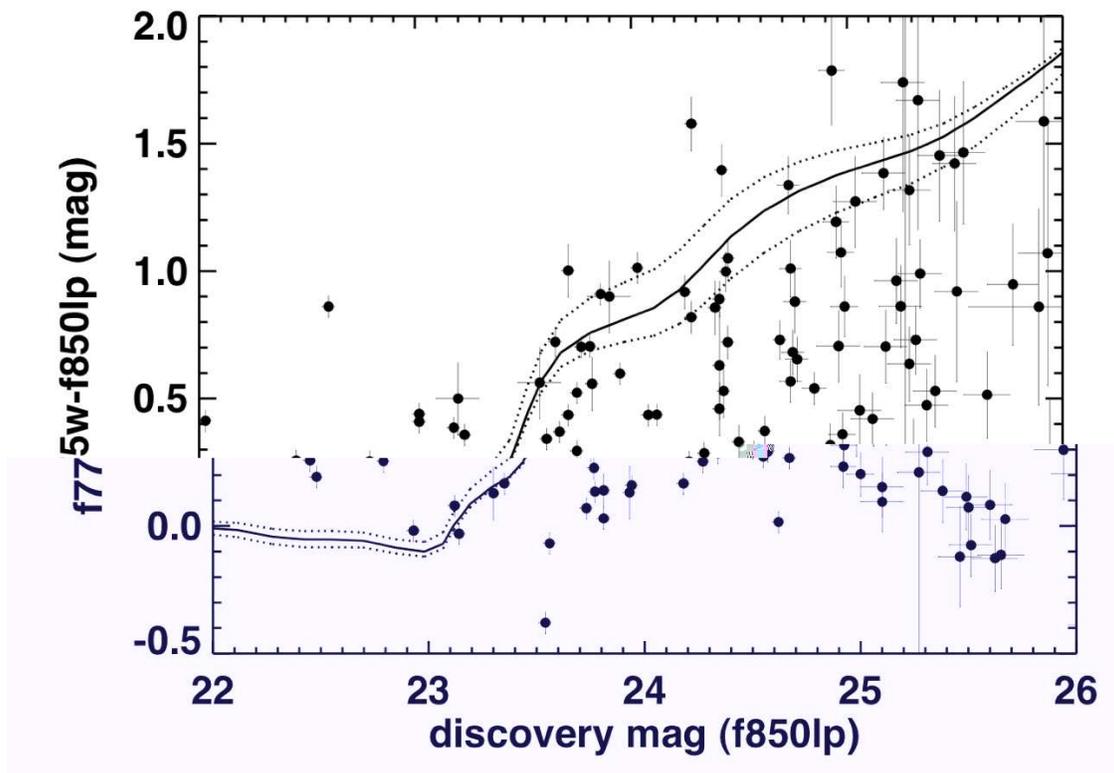}
\figcaption[f5.eps]{The (F775W-F850LP) color as a function of detection magnitudes in F850LP filter for the full sample of detected Type Ia and CC SNe. Solid line shows the color-magnitude relation for a Type Ia SN at peak magnitude (with the relation at +/- 3 days as dotted lines). We detect a number of faint SNe at $m>25$, but most of them are too blue to be distant Type Ia, instead, color (together with redshift and lightcurve shape) suggest that these are CC SNe at $z<1$. 
\label{fig5}}
\end{figure}

In Figure \ref{fig5}, we show the (F775W-F850LP) color as a function of detection magnitude in the F850LP filter for the full sample of SNe (Type Ia and CC). In the figure, we also plot the color-magnitude relation for a Type Ia SN observed near the peak on the lightcurve. The relation shows that as the Type Ia becomes more distant, it becomes both fainter and redder. The figure clearly illustrates that we are able to detect a number of SNe down to the detection limit at $m\sim 26$. However, the figure also shows that the majority of the faint objects with $m>25$~have colors F775W-F850LP$\lsim$0.8, which is too blue to be consistent with high redshift Type Ia SNe. Based on colors (in combination with redshift and lightcurve fitting), we find that these objects are almost exclusively CC SNe at $z<1$.

To evaluate how the derived rates depend on a correct determination of the detection efficiency, we recalculate our rates after assuming that we reach the 50\% detection efficiency 0.2 mag brighter than the limit derived in Strolger et al. (2004). As a test case, we also give rates assuming
a 0.5 mag brighter cutoff. The right panel of Figure \ref{fig4} shows how the derived rates change if we assume magnitude limits that are brighter than the nominal values. Black dots are nominal rates, while red and blue dots show the cases where we have assumed a brighter cutoff by 0.2 and 0.5 mag compared to $m_c$=25.94.

Since the highest redshift bin will contain mostly faint SNe, we expect the effect of changing the efficiency to be largest here, which is consistent with what the figure shows. The change in rate is, however, relatively small. Adopting a 0.2 mag brighter cutoff ($m_c$=25.74) only increases the rate by 13\%~in the highest redshift bin. Even with a detection limit 0.5 brighter ($m_c$=25.44), the trend of a strong decrease in the rates at $z>1.4$~remains. The investigation above shows that the drop in rate of SNe Ia in the high redshift bin is robust for viable detection efficiencies.

Additional support that we are not missing a number of high redshift SNe due to overestimating the detection efficiency comes from Kuznetsova et al. (2007). They independently searched the GOODS Cycle 11 data-set and four additional epochs in GOODS-N collected during Spring-Summer 2004 (which are also included in our investigation). No additional high redshift SNe were found by Kuznetsova et al., in fact, their $z>1.4$~sample includes the same three SNe as our sample. However, due to different selection criteria, only one of high redshift SNe is included in the sample used by Kuznetsova et al. to calculate the SNR. Figure \ref{fig1} shows that the rate calculated by Kuznetsova et al. (2007) decreases at high redshift similar to what we found here, although the error bars are larger so that the decrease is not as significant. Furhermore, Rodney \& Tonry (2007) have also searched the GOODS fields without finding any additional SN candidates at $z~>$ 1.     
\subsection{WD explosion efficiency}
Both the one-component and the two-component models rely on one strict assumption: that the WD explosion efficiency, i.e., the fraction of all WDs that explode as SN, stays at a fixed value at all redshifts. If this assumption is relaxed, the two-component model can be made consistent with a drop in the rate at $z>1.4$~by assuming that a decreasing fraction of WDs explode as Type Ia SNe at higher redshift.

The models by Kobayashi et al. (1998) suggest that the metallicity must reach [Fe/H]$\ge -1$~in order for the accretion onto the WD progenitor to be efficient enough to result in a Type Ia explosion. With the overall metallicity decreasing at higher redshift, Kobayashi et al. predicted a cutoff in the Type Ia explosion at $z>1.4$. Refined models by Nomoto et al. (2000) suggests a higher cutoff at $z\sim 2$~in spiral galaxies and at $z\sim 2.5$~in elliptical galaxies. Recent models by Kobayashi \& Nomoto (2008), based on the single degenerate scenario, suggest that the Type Ia rate should be low in spiral galaxies at $z\gsim 1$~due to the lower metallicity in these systems. This model leads to a scenario where the Type Ia rate peaks at $z\sim 1$~and thereafter declines, consistent with what our results suggests. 

\section{Are Type Ia Supernovae Bimodal?}
The new measurements of the Type Ia SNR to $z\sim1.6$~presented here are very similar to our previous results derived only from the first-year sample. Based solely from the dearth of events at $z>1.4$~in this complete sample, it would be difficult to conclude that there could be strong bimodality in the delay time distribution with a dominating prompt component. Instead, our investigation suggests that a major fraction of the Type Ia SNe should have a long characteristic delay time in the order of $\tau \sim 3$~Gyr. Therefore, our results are difficult to reconcile with the models of SB05 and N06, which find that $>$90\% of the Type Ia rate at $z>1.4$~should be prompt. In more agreement with our results is the bimodal model of Mannucci et al. (2006; 2007), which suggests that the old component contributes $\sim 50$\% to the Type Ia SNe events.

Until recently, only $HST$ surveys could provide insight on the redshift regime critical to this conclusion. The clear advantage has been the remarkable sensitivity and unparalleled resolution that $HST$ + ACS provides, thusly producing the deepest and most complete surveys for supernovae to date. However, recent ground-based ``ultra-deep'' surveys are beginning to close the gap in surveyable depth. The Subaru Deep Field (SDF) has collaterally produced a deep supernova survey that is comparable to GOODS (albeit with much poorer spatial resolution due to atmospheric effects and yet only a one-epoch search), with a $z'$-band limit of approximately $26.3$~mag (AB). The results from their initial sample (Poznanski et al. 2007) show a appreciable similarity to the GOODS first-year sample (when corrected for differences in effective volume and control time of the surveys) in the redshift distribution of SNe~Ia, including the clear reduction in the number of SNe~Ia above $z>1.2$. Their analysis of SDF survey detection efficiency has likewise concluded that this reduction cannot be explained as simply a loss in sensitivity to SNe~Ia at these redshifts.

Still, there are notable differences in the rates derived from these similar samples. Foremost, Poznanski et al. (2007)  measure of an appreciably higher (factor $\sim$2) Type Ia SNR in the highest redshift bins (at $z> 1.4$) than is presented in this manuscript. However, of the three Type Ia SNe in the highest redshift bin in Poznanski et al., two have large uncertainties in both type determination and redshift. This is reflected in the large error-bars given by Poznanski et al., which makes their rate consistent with the rate found here. But despite the uncertainty of these few supernovae, it is encouraging the SDF supernova survey will continue to gather a more definitive measure of the SNR at $z>1.4$~less affected by these individual uncertainties. We conclude that at present, the observed SN rates at high redshift can not alone rule out a scenario where the delay time distribution has a large contribution from an old ($\tau \sim 3$~Gyr) population.

There are, however, a number of results that show strong support for a bimodal model with a significant fraction of the Type Ia events coming from a prompt component, as discussed in \S 1. It is therefore desirable to try to reconcile the observed rate with the bimodal model. As shown in \S 4 and Figure \ref{fig2}, the models by M07 gives a better fit to the data compared to the other bimodal models. There are two main reasons for this better fit. First, M07 assumes a significantly larger fraction from the old Type Ia progenitor component compared to the other models and second, M07 use different assumptions about the cosmic SFR and dust extinction. When we in \S 4 calculated the best-fitting one-component model for our data, we used the same assumptions about dust extinction in the population of galaxies that produced the UV-luminosity (from which the SFR was calculated) and that produced the SNe. In M07, a large fraction of the SFR is assumed to be hidden in dusty environments and will be missed in optical searches (we refer to this as the IR part of the SFR). The total SFR used by these authors therefore has a different shape compared to SFRs based on UV luminosities only (UV SFR). At the same time, most of the SNe occurring in these obscured galaxies will also be missed from SN searches. In order to compare our results with this model, we use the SFR in M07, taking into account the missing SFR as well as the expected fraction of missed Type Ia SNe as given in M07. Making a fit to the data, we find again that the narrow Gaussian model gives the best fit, but now with a characteristic delay time of $\tau~\sim~2.9$~Gyr, compared to $\tau~\sim~3.4$~Gyr for our original model. For a wide Gaussian delay time model, our model gives $\tau~\sim~3.1$~Gyr, while the M07 model gives $\tau~\sim~2.0$~Gyr. Investigating these differences in some detail, we find that part of the differences is due to the 'hidden' IR SFR, but also that a significant fraction of the change is due to different assumptions on the shape of the UV SFR. If we use the IR SFR given by M07 in combination with the UV SFR adopted in this paper (taken from Giavalisco et al. 2004), we find that even though we do include the missing fraction of the SFR and the SNR, the delay times do not change considerably. For the best-fitting narrow Gaussian model we find  $\tau~\sim~3.3$~Gyr (compared to our original result  $\tau~\sim~3.4$~Gyr), while for the wide Gaussian we find  $\tau~\sim~2.6$~Gyr ($\tau~\sim~3.1$~Gyr). This highlights an additional problem for deriving the delay time function, as long as there are significant uncertainties in the evolution of the cosmic SFR, the derivation of the delay time function will be hampered.

We find that taking missing fraction of SFR and SNR into account, according to the M07 model, leads to a slightly lower characteristic delay time, but that this result is somewhat dependent of the assumption of the shape of the SFR. The characteristic delay-time we find is in the order of a few Gyr, implying that a significant fraction of the SNR should be delayed and therefore decoupled from the ongoing SFR. 
We also note that in a recent paper, Botticella et al. (2008) studied the evolution of both core collapse and Type Ia SNe to $z\sim$~0.2--0.3. From the stronger evolution in the core collapse SN rate they found, they conclude that a significant fraction of the Type Ia progenitors has a delay time 2--3 Gyr, consistent with what found here. 
\section{Comparisons at lower redshift}
Besides at high redshift, there are other notable differences between the GOODS and other results. In the rate measurement within the two bins spanning the range of approximately $0.4\la z\la1.0$, Poznanski et al. (2007) have found roughly half the Type Ia supernovae near $z\sim1$~as was seen in the GOODS, and essentially none at lower-$z$. Given the single search epoch and acute solid angle surveyed, it is not unlikely for the SDF survey to have a null result at lower redshifts. However, the differences near $z\sim0.8$~seem appreciable. It serves as another example of a growing problem; the relatively large discrepancies in rate measurements by several authors in this redshift range (see Figure \ref{fig1}). It is disappointing that there does not as yet appear to be a resolution in these measured discrepancies, as there could be great power in combining rate results to gain further insight on the modality of the SN~Ia delay-time distributions, and ultimately the progenitors mechanisms of Type Ia supernovae. But until then, this analysis will remain necessarily limited.

\section{Conclusions and summary}
Here we present new measurements of the Type Ia SNR to $z\sim 1.6$. Similar to our previous results based on a smaller sample, these observations show a decrease in the SNR at redshifts $z\gsim 1.4$, with a high significance. The results are consistent with a characteristic delay time in the order of $\tau=2-3$~Gyr. Recent two-component models for the Type Ia SNR, with one dominating prompt and one less prominent delayed channel, seems to fit low redshift SNR data well. These models are also consistent with the findings that the SNR is higher in galaxies with a higher star formation, and they may also explain the Fe content of the inter-cluster medium in clusters of galaxies. However, these two-component models predicts rates at $z>1.4$~that deviates from the measured rates from this investigation. Here we have discussed possible solutions to this discrepancy and found:
\begin{itemize}   
\item{It is unlikely that the difference between model predicted rates and observed rates is due to statistical fluctuations or cosmic variance.}
\item{It is also unlikely that the low rate we measure is due to an underestimate of the host galaxy dust extinction or an overestimate detection efficiency.}
\item{A bimodal model with a larger fraction of delayed Type Ia and that takes into account SFR hidden by dust results in a better fit to data.}
\item{Another possible scenario that would result in a decrease in the high redshift SNR is if the WD explosion efficiency deceases at high redshift.}
\end{itemize}
\acknowledgments{
We thank the anonymous referee for valuable comments and suggestions. Based on observations made with the NASA/ESA Hubble Space Telescope, obtained at the Space Telescope Science Institute, which is operated by the Association of Universities for Research in Astronomy, Inc., under NASA contract NAS 5-26555. These observations are associated with programs GO-9352, GO-9425, GO-9583, GO-9728, GO-10189, GO-10339, and GO-10340.
}


\clearpage



\clearpage
\begin{deluxetable}{lcccccc}
\tabletypesize{\scriptsize}
\tablewidth{0pt}
\tablecaption{Type Is Supernova Rates}
\tablecolumns{7}
\tablehead{
\colhead{}Redshift bin & Effective redshift & SNR & error(stat) & error(syst) & N & N$_{dist}$
}
\startdata
$0.2< z\le0.6$ & 0.47 & 0.80 & $^{+0.37}_{-0.27}$ & $^{+1.66}_{-0.26}$ & 8 & 8.8 \\
$0.6< z\le1.0$ & 0.83 & 1.30 & $^{+0.33}_{-0.27}$ & $^{+0.73}_{-0.51}$ & 25 & 23.5\\
$1.0< z\le1.4$ & 1.21 & 1.32 & $^{+0.36}_{-0.29}$ & $^{+0.38}_{-0.32}$ & 20 & 20.2 \\
$1.4< z\le1.8$ & 1.61 & 0.42 & $^{+0.39}_{-0.23}$ & $^{+0.19}_{-0.14}$ & 3 & 3.1
\enddata
\tablecomments{The effective redshift is defined as the redshift that divides the volume in the redshift bin into equal halves. Rates are given in units $yr^{-1}~Mpc^{-3}~10^{-4}~h_{70}^3$, assuming a cosmology with $\Omega_M=0.3$ and $\Omega_{\Lambda}=0.7$. First quoted errors are statistical and represent 68.3\% confidence intervals. Second errors are systematic and include the possibility that all SNe with uncertainty in type determination (i.e., bronze type) have been misclassified, as well as other possible sources discussed in Dahlen et al. (2004). Note that the latter errors are non-Gaussian. Last two columns give the number of SNe in each bin, where N is the 'raw' counts and N$_{dist}$~gives the number after taking into account the redshift probability distribution of the SNe.
}
\label{tab1}
\end{deluxetable}

\end{document}